\documentstyle[aps]{revtex}

\begin{document}
\draft
\title{Density probability distribution in one-dimensional
  polytropic gas dynamics}
\author{Thierry Passot}
\address{Observatoire de la C\^ote d'Azur, B.P.\ 4229, 06304, Nice
  Cedex 4, France}
\author{Enrique V\'azquez-Semadeni}
\address{Instituto de Astronom\'\i a, UNAM, Apdo. Postal 70-264,
  M\'exico, D.\ F.\ 04510, M\'exico} 
\date{\today}
\maketitle
\begin{abstract}

We discuss the generation and statistics of the density fluctuations
in highly compressible polytropic turbulence, based on a simple model
and one-dimensional numerical simulations.
Observing that density structures tend to form
in a  hierarchical manner, we assume that density
fluctuations follow a random multiplicative process. When the
polytropic exponent $\gamma$ is equal to 
unity, the local Mach number is independent of the
density, and our assumption leads us to expect that the probability
density function (PDF) of the density field is a lognormal. 
This isothermal case is found to be
singular, with a dispersion $\sigma_s^2$ which scales like the
square turbulent Mach number $\tilde M^2$, where $s\equiv \ln \rho$ 
and $\rho$ is the fluid density. This leads to
much higher fluctuations than those due to shock jump relations. 

Extrapolating the model to the case
$\gamma \not =1$, we find that, as the Mach number becomes large, 
the density PDF is expected to
asymptotically approach a power-law regime, at high densities when
$\gamma<1$, and at low densities when $\gamma>1$. This effect can be
traced back to the fact that the pressure term in the momentum
equation varies exponentially with $s$, thus 
opposing the growth of fluctuations on one side of the PDF, while being
negligible on the other side. This also causes the dispersion
$\sigma_s^2$ to grow more slowly than $\tilde M^2$ when $\gamma\not=1$. In
view of these results, we suggest that Burgers flow is a 
singular case not approached by the high-$\tilde M$ 
limit, with a PDF that develops power laws on
both sides.

\end{abstract}
\pacs{47.27.Ak, 47.40.Ki, 95.30.Lz}

\section{Introduction}\label{introduction}

The formation of density structures by the velocity
field of highly compressible turbulence is of great
interest in astrophysics. The determination of their typical  
amplitude, size and volume filling factor 
poses significant difficulties since it requires a knowledge of the
full statistics. As a first step, we shall concentrate in this paper
on one-point statistics and more specifically on the probability
density function (PDF)  of the density fluctuations  in
one-dimensional (1D) turbulent flows.

It is well known that the density jump in a shock depends directly on
the cooling ability of the fluid. Thus, for an adiabatic flow the
maximum density jump is 4, for an isothermal flow it is $\sim
M_a^2$ \cite{landau}, and for nearly isobaric flows it is $\sim
e^{M_a^2}$ \cite{VPP96}, where $M_a$ is the Mach number ahead of the
shock. The net cooling ability of a flow can be 
conveniently parameterized by the polytropic exponent $\gamma$, so
that the thermal pressure $P$ is given by $P=K\rho^\gamma$,
where $\rho$ is the fluid density \cite{footnote}.
Isothermal flows have $\gamma=1$, and isobaric
flows have $\gamma=0$. Note that $\gamma<0$ corresponds to the
isobaric mode of the thermal instability (see, e.g.,
\cite{B95}). Thus, in general, the amplitude of 
the turbulent density fluctuations will be a function of $\gamma$.

Previous work with isothermal flows had
suggested that the PDF is log-normal \cite{V94,PNJ97}, while for
Burgers flows a 
power-law PDF has been reported \cite{GK93}. More recently, evidence
that flows with effective polytropic indices $0<\gamma<1$ also develop
power-law tails at high densities has been presented \cite{SVCP97}. 
In order to resolve this discrepancy, we present a series of
1D numerical simulations of polytropic gas 
turbulence with random forcing, in which the polytropic exponent $\gamma$
parameterizes the compressibility of the flow. 
We have chosen to use 1D simulations in order to perform a large
number of experiments at a sufficiently high resolution, integrated
over very long time intervals, allowing us to collect large
statistical samples.

The simulations have
three governing parameters: the polytropic index $\gamma$, the Mach
number $M$, and the 
Reynolds number $R$. We keep the Reynolds number fixed, and explore
the effects of varying $\gamma$ and $M$ on the resulting density
PDF. We find that varying these two parameters is not 
equivalent. Variation of $\gamma$ induces a clear qualitative
variation of the density PDF, which, at large Mach number, displays
a power-law tail at high 
densities for $0<\gamma<1$, becomes log-normal at $\gamma=1$, and 
develops a power-law tail at low densities
for $\gamma>1$. This suggests a symmetry about the case $\gamma=1$, which
we also explore. Variation of the Mach number, on the other
hand, only appears to 
induce a quantitative change, in such a way that increasing $M$ 
augments the width of the PDF.

The plan of the paper is as follows. In sec.\ \ref{numerics} we
describe the equations solved and the numerical method. In sec.\ 
\ref{statistics} we describe the statistics of the various fields, in
terms of their PDFs, together with a tentative model and a discussion
of the Burgers case.
Section \ref{conclusions} is devoted to a  discussion on the choice
of the forcing, together with a summary of our results.

\section{Equations and numerical method}\label{numerics}

We choose to concentrate on one-dimensional forced polytropic gas
dynamics, governed by the following non-dimensionalized equations
\begin{eqnarray}
&& \partial_t u+ u\partial_x u=-\frac{1}{\gamma M^2}\frac{\partial_x
  \rho^{\gamma}}{\rho} +\frac{1}{R} \partial_{xx} u +a \label{eq:baseu}\\
&& \partial_t \rho +\partial_x (\rho u) =0 \label{eq:baser}
\end{eqnarray}
where $u$ is the velocity of the fluid in units of $U$,  $\rho$ the
density in units of $\rho_0$, $\gamma$ 
the polytropic index and $M$ the Mach number of the unit 
velocity $U$ at the unit density $\rho_0$. The equations are driven by an
acceleration $a$ with zero mean. The  Reynolds number is
$R=\frac{UL}{\nu}$, where $L$ is the size of the
domain and $\nu$ the kinematic viscosity
chosen constant to ensure the conservation  of the mean velocity 
$\langle u\rangle=\frac{1}{L}\int u dx$. The viscous term is kept as
small as possible and is only here to prevent numerical
blow-up. Note that the
``correct'' form of the viscous term is obtained after replacing $\nu$
by the ratio $\mu/\rho$, where the dynamical viscosity $\mu$ is
usually considered independent of the density.  
The equations then conserve the
momentum $\int \rho u dx$ if the acceleration $a$ in
eq. (\ref{eq:baseu})  is also replaced by the ratio of a force $f$ to
the density $\rho$.  
The dynamics that results in this case is very different due to the
dependence of the driving term with respect to the density, as
discussed in the last Section.

For large Mach number simulations, it was found necessary to 
smooth density gradients, using a mass
diffusion term of the form $\mu_r \partial_{xx}\rho$ in the
right-hand side of eq. (\ref{eq:baser}). Total mass is still conserved
in the presence of this term, and if $\mu_r$ is taken sufficiently
small, it has been tested that it does not affect the dynamics in a
way that could modify our conclusions. 

We also found it convenient to solve
eqs. (\ref{eq:baseu})-(\ref{eq:baser})  using the variable
$s\equiv \ln\rho$. The numerical code uses a standard pseudo-spectral
method with periodic boundary conditions. Time advance is performed
using a Crank-Nicholson scheme for the linear terms and an
Adams-Bashforth scheme for the nonlinear ones.
For all the runs presented in this paper, the kinematic viscosity has
been fixed to $\nu= 3\times 10^{-3}$. For runs with $M\ge 3$, we have
$\mu_r=5\times 10^{-4}$.

The acceleration $a$ is prescribed in Fourier space. Its
spectrum has a constant amplitude (equal to 0.6) on wavenumbers
$1\le k \le 19$ and  phases chosen randomly with a correlation time $t_{\rm
  cor}=0.003$. Resolution ranges from $N=3072$ to $N=6144$ grid points
for the runs with $M\ge 6$.

We perform one point statistics of the simulations, both for the
density and the velocity derivative,
keeping the forcing and the viscosity constant.
All simulations start with zero initial velocity and constant
density.

In order 
to obtain reasonably sampled histograms of the one-dimensional
fields, which contain only $N$ spatial data points, we sum the
histograms over time, sampling at intervals of $0.1$ time units,
integrating over a total of 150 time units. However, we have found that,
since the simulations start with uniform density, the first several
samples must be discarded, since they bias the density histogram 
near $\rho=1$. We typically skip the first 20 temporal
samples (2 time units). The PDFs thus computed contain
roughly 4  million data points.
Note that longer integration times are needed at larger Mach number in
order to reach a statistically relevant sample, the
sound crossing time of the integration domain being larger as $M$
increases.


\section{A model for the density PDF}\label{statistics}

\subsection{Properties of the governing equations}

Before describing our model for the density PDF, it is instructive to
rewrite the governing equations in the inviscid, unforced case, using the
variable $v =(1-\gamma) \ln \rho$ when $\gamma \ne 1$ and
$s=\ln \rho$ when $\gamma =1$.
We get, for  $\gamma \ne 1$
\begin{eqnarray}
&&\frac{Du}{Dt} =\frac{1}{(1-\gamma)M^2}\frac{\partial}{\partial
  x}e^{-v}\label {eq:Sigu} \\
&& \frac{Dv}{Dt} = -(1-\gamma) \frac{\partial}{\partial  x} u
\label{eq:Sigr} 
\end{eqnarray}
and for $\gamma=1$

\begin{eqnarray}
&&\frac{Du}{Dt} =-\frac{1}{M^2}\frac{\partial}{\partial
  x}s \label{eq:Su} \\
&& \frac{D s}{Dt} = - \frac{\partial}{\partial  x} u  \label{eq:Sr}
\end{eqnarray}
where $\frac{D}{Dt}$ stands for the convective derivative.
The variable $v$ is, up to an additive constant, the logarithm of
the square of the sound speed, and when $\gamma=1$, becomes identically zero.

These equations can be rewritten in Riemann invariant
form. For $\gamma\ne 1$, they read
\begin{equation} \label{eq:RI}
\bigl[\partial_t +(u\pm c)\partial_x\bigr](u\pm
\frac{2c}{(\gamma-1)})=0,
\end{equation}
where  $c={\rho^{\frac{\gamma-1}{2}}}/{M}$ is the sound speed, while 
in the singular case $\gamma=1$ these equations become
\begin{equation}\label{eq:RIgam1}
\bigl[\partial_t +(u\pm \frac{1}{M})\partial_x\bigr](u\pm
\frac{\ln \rho }{M})=0.
\end{equation}

A number of interesting remarks can be made on the previous
equations. 

(i) When $\gamma=1$, eqs. (\ref{eq:Su})-(\ref{eq:Sr}) are invariant
upon the change 
$s\rightarrow s+b$, where $b$ is an arbitrary constant.
Indeed, the sound speed does not depend on
the local density of the fluid. 

(ii) In the general case, if we substitute $\gamma$ by $2-\gamma$ and
$\rho$ by $1/\rho$, we observe that the Riemann invariants $z^{\pm}=u\pm
\frac{2c}{(\gamma-1)}$ are
exchanged, while their speeds $u\pm c$ remain unchanged. 
We shall now explore the implications of this remark on the statistics
of the density fluctuations in the weakly compressible regime.
For small values of the Mach number, a reductive perturbation
expansion can be performed on the viscous equations and it has been
shown \cite{T76} (see also \cite{TT73})
that one-dimensional  compressible turbulence 
reduces essentially to the superposition of the solutions of two
Burgers equations describing nonlinear wave 
propagation in opposite directions.
More precisely (considering eqs. (\ref{eq:baseu}-\ref{eq:baser}) with
$M=1$), if we denote by 
$\rho'$ and $u'$ the perturbations of 
the basic state $(\rho=1, u=0)$, Tokunaga obtained (\cite{T76})
\begin{eqnarray}
&&\rho'=\frac {2\epsilon}{\gamma+1}( F_1(\xi_1,
\tau)-F_2(\xi_2,\tau)) \label{eq:Tok1} \\ 
&& u'=\frac {2\epsilon}{\gamma+1}( F_1(\xi_1,
\tau)+F_2(\xi_2,\tau)) \label{eq:Tok2}
\end{eqnarray}
where $\epsilon$ is the order of magnitude of the nonlinear waves.
The new coordinates $\xi_i$ and $\tau$ are defined by
\begin{eqnarray}
&&\xi_i=\epsilon \bigl[ x -r_i t -\phi_i (x,t)\bigr]\\
&& \tau=\epsilon^2 t
\end{eqnarray}
where $r_1=1$ and $r_2=-1$ and the phase functions obey
\begin{eqnarray}
\phi_1=\frac{1}{2}\frac{3-\gamma}{1+\gamma}\int^{\xi_2}
F_2(\xi,\tau)d\xi +\theta_1 \\
\phi_2=-\frac{1}{2}\frac{3-\gamma}{1+\gamma}\int^{\xi_1}
F_1(\xi,\tau)d\xi +\theta_2,
\end{eqnarray}
with $\theta_i$ arbitrary constants determined by the initial
conditions.
Finally the functions $F_i$ (simply related to the Riemann invariants
$z^{\pm}$) satisfy the Burgers equations
\begin{equation}
\partial_{\tau} F_i +F_i\partial_{\xi_i} F_i=\frac{\nu}{2}
\partial_{\xi_i \xi_i} F_i.
\end{equation}
 The fields $F_i$ evolve 
almost independently, with the same dynamical equation, except for
phase shifts, a higher order effect most important during collisions of shock waves. 
Given some initial conditions for $\rho'$ and $u'$, the substitution
$\rho\rightarrow 1/\rho$ (or $\rho' \rightarrow -\rho'$), 
and $\gamma\rightarrow 2-\gamma$ leads
to the replacement of $F_1$ and $F_2$ by $F_2(3-\gamma)/(1+\gamma)$
and $F_1(3-\gamma)/(1+\gamma)$ respectively.
For a vanishingly small viscosity $\nu$, the rescaling of the amplitudes 
$F_1$ and $F_2$ can be absorbed in a rescaling of the variables $\xi_i$.
Except for this stretching of the space and time variables, this substitution
will thus lead to the same fluctuations occurring at different locations.
As a consequence, we can expect that the probability density 
functions of the cases $\gamma$ and $2-\gamma$ for small
values of $M$ will be closely related after the change $\rho\rightarrow
1/\rho$.
The case of higher Mach numbers is more delicate due to the additional
problem of mass conservation, rendering impossible a symmetry
between $\rho$ and $1/\rho$. This question is addressed below.

(iii) The substitution $\gamma \rightarrow 2-\gamma$ can also be examined at
the level of eqs. (\ref{eq:Sigu})-(\ref{eq:Sigr}). Its effect is simply
to change the sign of the right-hand sides. For $\gamma <1$,
eq. (\ref{eq:Sigr}) shows that positive values of $v $ (in this
case associated with  density peaks)  are mostly
created by shocks (associated with negative velocity gradients). Looking at 
eq. (\ref{eq:Sigu}), we see that as $v$ increases, the pressure term
becomes exponentially small and thus cannot prevent
the formation of very strong peaks. Negative values of
$v$ (here associated with density voids) are created by expansion
waves, but in that case the pressure 
increases exponentially with decreasing values of $v$ leading to a
rapid saturation of this process. As a consequence, we
expect that for $\gamma <1$ the PDF of $v$ will be 
significantly more populated at positive rather than at negative
values. For $\gamma$ larger than unity the PDF of $v$ will be
similar, the formation of
positive values of $v$ (now associated with density voids) being still
unhindered by the pressure.
It results that the PDF of $s=\ln\rho$ for $\gamma >1$ will appear
similar to that for $\gamma <1$ after we change $s\rightarrow -s$. 
  
(iv) When $\gamma=1$, the behavior is very different since the acceleration
due to the pressure term is simply proportional to $-\partial_x s$ and
thus never becomes negligible. We  expect a symmetry in the PDF
of $s$, positive and negative values of $s$ being equally created, by
shocks and expansion waves respectively. 

(v) Finally, it is also useful to discuss the shock jump relations
for the polytropic equations. Denoting by $X$ the ratio of the
post-shock to the pre-shock density, we have (see \cite{VPP96})
\begin{equation}
X^{1+\gamma}-(1+\gamma m^2)X +\gamma m^2 =0 \label {eq:Jump}
\end{equation}
where $m$ is the upstream Mach number in the reference frame of the
shock. 
This equation shows that for $\gamma=1$, $X=m^2$ and that the jump $X$
increases more slowly than $m^2$ for $\gamma>1$, while it increases faster
than  $m^2$ for $\gamma<1$ with, as $\gamma \rightarrow 0$, $X\sim
e^{m^2}$. 
For weak shocks, we get $X\approx 1+(m^2-1)\frac{2}{1+\gamma}$.
In this case, the shock velocity being close to the sound speed, we
can write $m=1+\frac{u}{2c}$, where $u$ is the velocity in the
simulation frame, leading to $X=1+\frac{\Delta
  \rho}{\rho}=1+\frac{u}{c}\frac{2}{1+\gamma}$. We thus get $\frac{\Delta
  \rho}{\rho}\sim  m_s \frac{2}{1+\gamma}$, with $m_s=u/c$ 
denoting the Mach number in the simulation frame. 

Note that typical pressure fluctuations
created by almost incompressible turbulence scale like $\tilde M^2$ 
where the turbulent Mach number is defined as $\tilde M= u_{\rm
  rms}/c$ (here $u_{rms}$ is mostly made of solenoidal motions unlike
in our 1D simulations where it stands for purely compressible modes).
 This scaling corresponds to a balance between the 
pressure gradient and the nonlinear term. If entropy
fluctuations are not allowed (like with a polytropic state law), the
resulting density fluctuations also have 
to scale as $\tilde M^2$. In thermally forced turbulence however,
a Boussinesq-like balance obtains between temperature and density
fluctuations, maintaining pressure fluctuations of order
$\tilde M^2$, while allowing for much larger values of density and
temperature fluctuations \cite{BLP92}.

In weakly nonlinear acoustics, the pressure term is balanced by
the velocity time derivative and we recover the scaling  $\frac{\delta
  \rho}{\rho}\sim \tilde M$ obtained for weak individual shocks.

\subsection {The case $\gamma=1$}

The main idea of our model is that density fluctuations are built up in
a hierarchical process \cite{V94}. After a shock (respectively an
expansion wave) 
passes through a given region of 
mean density $\rho_0$, the density reaches a new value $\rho_1$,
larger (respectively smaller) than $\rho_0$.  In this
region new fluctuations can be created, changing the local
value $\rho_1$ to $\rho_2$ and so on.  Of course the
dynamical equations constrain this process. For example,
due to mass conservation, arbitrarily high values of the density
can only be reached in very localized and thin peaks. We thus expect
this hierarchical process to saturate at some value $s_{+}>0$.
A similar saturation should occur for low densities at some value
$s_{-}<0$, with probably $|s_{-}|>|s_{+}|$, due to the fact that larger
voids can be created without violating the mass conservation constraint.

The build up of these density
fluctuations is a random multiplicative  process which, at the level
of the variable $s$, becomes additive. The random variable
$s$ is thus the sum of individual random variables (the density
fluctuations), each having the same
probability distribution. The latter fact follows from the invariance
of the equations at $\gamma=1$ under the change $s\rightarrow s+s_0$,
which furthermore implies that each individual jump has the same
average magnitude, related to the Mach number of the flow but
independent of the local density.  
The sum of identical random processes is known to have a Gaussian
distribution, due to the Central Limit Theorem, whatever the
distribution of the individual processes. The PDF of $s$ is thus
expected to follow a normal distribution. 

The variance of the random variable $s$ can be estimated using 
the size of the typical fluctuations associated  both to shocks and
expansion waves. The case of shock waves has been discussed above. 
At small values of $M$, $\rho_{i+1}/\rho_i \sim 1+\tilde M$
(with the Mach number $\tilde M= u_{\rm rms}/c$ )
so that $\delta s =\ln(\rho_{i+1}/\rho_i)=\ln (1+\delta\rho/\rho)\approx \ln (1+\tilde M)
\sim\tilde M$. At  high Mach numbers, the individual jumps obey $\Delta s \sim \ln
\tilde M^2$. 
For expansion waves, the balance of the time derivative of $s$ and of
 the positive velocity gradient in eq. (\ref{eq:Sr}), gives $s\sim
M$, regardless of the value of $M$ since the term $us_x$ is
smaller than the term $s_t$ if the density decreases uniformly in
space (note that the decrease of $\rho$ is exponential in time). 
Thus, with this mechanism,  the density  decreases in the
center of the expansion waves while it increases on the edges, until
pressure blocks the process. In the case $\gamma=1$ pressure acts
symmetrically in $s$ and we thus get positive and negative
fluctuations which are of the same order of magnitude, 
and much larger than those due to shocks.
We thus expect that $\sigma_s\sim \tilde M$ for a large range of
values of the Mach number.

 From the previous dicussion we can expect the PDF of the variable $s$
to be given by
\begin{equation}
P(s)ds=\frac{1}{\sqrt{2\pi \sigma_s^2}}\exp\bigl(-\frac{(s-s_o)^2}
{2\sigma_s^2}\bigr) ds
\end{equation}
with $\sigma_s^2=\beta \tilde M^2$, and $\beta$ a proportionality
constant. The maximum of this distribution $s_o$ is simply related to
$\sigma_s$ due to the constraint of mass conservation.
Writing $\langle\rho\rangle=\int_{-\infty}^{+\infty} e^s P(s)
ds =1$ we find $s_o = -\frac{1}{2}\sigma_s^2$ (see below).
Note that the PDF of $\rho$ is related to that of $s$ by
$P_{\rho}(\rho)=P(\ln\rho)/\rho$.

The predictions of this model can be tested against results from numerical
simulations. Figure \ref{sig_vs_M_g1} (top panel) shows a plot of $\log
(\sigma_s)$ vs.\ 
$\log(\tilde M)$ obtained by combining data from several simulations with
$M=0.5$, 1, 2, 3, 4.5, 6 and 10. These data were obtained by
computing $\tilde M$ 
and $\sigma_s$ for the accumulated density and velocity fields over 
100 subsequent outputs of the simulations (spanning a total duration
of 10 time units) for each point in Fig.\ \ref{sig_vs_M_g1}.
This plot shows that $\sigma_s^2\approx \beta \tilde M^2$, with $\beta
\approx 1$, with a very good accuracy, up to the highest Mach numbers
reached in our simulations. On the other hand, we see in the bottom
panel of the same figure,
which displays $\log\sigma_{\rho}$ vs. $\log \tilde M$, that the density
standard deviation also scales like $\tilde M$ for small values of
$\tilde M$, while for $\tilde M \gtrsim 0.5$ the points curve up, a
reflection of 
the relation $\sigma_{\rho}^2=e^{\sigma_s^2}-1$ between the two variances
when $\rho$ obeys a log-normal distribution. The relation $s_o =
-\frac{1}{2}\sigma_s^2$ is also well verified 
numerically as can be seen form  Fig.\ \ref{s_vs_sigs_g1}.

We now display in Fig.\ \ref{pdfs_vs_M} the logarithm of the $s$-histograms
for three runs with $\gamma=1$ and $M=0.5$, 2, and 6. 
Fits with parabolas are shown in dashed lines and show that, to a very
good approximation, the PDFs of the density are in all three cases
log-normals. An estimation of the widths and maxima of these
distributions also shows a very good agreement with the predictions
$\sigma_s\approx M$ and $s_o =-0.5 \sigma_s^2$.

The distribution of the velocity derivative $u_x$ is shown in
Fig.\ \ref{pdf_ux} for $\gamma=1$ and $M=6$. This distribution is found to be
almost independent of the Mach number. It presents a long exponential
tail for negative values of $u_x$ and a strong drop off for large
values, analogous to the one found in the Burgers case \cite{GK93}.

\subsection{The case $\gamma\ne 1$ }

The difference between the case $\gamma=1$ and the cases
$\gamma\ne 1$ lies in the behavior of the pressure term as a function
of the local density of the fluid, an effect which is most visible
after comparing eq. (\ref{eq:Sigu}) with eq. (\ref{eq:Su}).
With the density-dependent rescaling $M\rightarrow  M(s;\gamma)=
Me^{\frac{1-\gamma}{2}s}$, the two equations identify, which only
means that we expect the small-fluctuation 
behavior of the case $\gamma\ne 1$ to be
identical to that of the case $\gamma=1$ in regions  where the local
logarithm of the density is close to $s$, when $\hat M(s;\gamma)$ is
substituted for $M$.  

The argument at the origin of the PDF of $s$ in the isothermal case is
based on the fact that the local Mach number of the flow is
independent of the local density. When $\gamma\ne 1$ this property
is violated and there is no reason to expect a log-normal PDF for the
density. We nevertheless  propose a heuristic  model, reproducing
most of the features of the PDF's obtained in our simulations, which
consists in taking the same functional form of the PDF as in the
isothermal case, but replacing $\tilde M$ by $ \hat M(s;\gamma)$, where 
$ \hat M(s;\gamma)$ now stands for the ``effective'' 
r.m.s.\ Mach number at the value $s$. This ``effective''
r.m.s.\ Mach number is defined as $ \tilde
M(s;\gamma)=u_{rms}/c(s)$, the local  turbulent Mach number, when
$s_{-}<s<s_{+}$, and by the constant $\tilde M(s_{+};\gamma)$ (respectively
$\tilde M(s_{-};\gamma)$ ) for $s>s_{+}$ (respectively $s<s_{-}$). These
cutoffs, which, as we shall see, are necessary for convergence, are
also physically meaningful, since the probability of new fluctuations
arising within previous peaks or voids decreases as the amplitudes of
the latter become larger because the fraction 
of space they occupy decreases.
The fact that the
cutoff occurs at larger values of $|s|$ for $s<0$ than for $s>0$ is
due to the larger filling factors of low density regions (see Fig.\
\ref{rho_fields}a and Fig.\ \ref{rho_fields}b for comparison).
A numerical check of this saturation property is possible if one
computes the scatter plot of the standard deviation for $s$ vs. the
mean value of $s$ in subregions of the integration domain for each snapshot.
Figure \ref{sig_vs_avg} shows these plots for $M=6$, $\gamma=0.5$ and
$\gamma=1.5$ in subregions of length $N/3$. A clear trend is visible,
indicating the change of the 
local Mach number with the local mean density. Moreover, we
clearly see that the saturation level for $s<0$ at $\gamma=1.5$
occurs at a much higher value of the Mach number than for $s>0$,
$\gamma=0.5$. 
Plots of $\sigma_s$ and $\sigma_{\rho}$ vs.\  $\tilde M$ for
$\gamma=0.5$ and $\gamma=1.5$ are also presented in Figs.\
\ref{sig_vs_M_g05} and \ref{sig_vs_M_g15}. 
They show that $\sigma_s$ increases more slowly than linearly
with $\tilde M$ for high Mach numbers. This results from the
asymmetry in the fluctuations of $s$ for $\gamma\ne 1$. While for
$\gamma=1$ the typical excursions of $s$ are of the order of $\tilde
M$ both for positive and negative values of $s$, when $\gamma>1$ for
example, pressure blocks the negative fluctuations of $s$ while
still allowing for fluctuations of order $\tilde M$ on the positive
side. The resulting variance $\sigma_s$ is thus expected to be smaller
than $\tilde M$. The same argument applies for $\gamma>1$ but then
fluctuations are of smaller magnitude when $s>0$.
Looking at the plot of $\sigma_{\rho}$  we see opposite trends for
$\gamma >1$ and $\gamma <1$. First, note that we do not expect the specific
relation  mentioned above between the variances of $s$ and $\rho$,
since the  distribution of $s$ is not Gaussian. Second, this trend is
easily interpreted if we recall that for $\gamma<1$ the density
fluctuations are in high peaks, while for $\gamma>1$ they consists of
large voids. In the former case the variance of $\rho$ can increase
greatly when $M$ is large, while in the latter case, the voids do not
contribute much in the variance, leading to a slower
increase of $\sigma_{\rho}$ with $\tilde M$.

The PDF will thus read
\begin{equation}
P(s;\gamma)ds=C(\gamma)\exp\Bigl[-\frac{s^2}{2\hat
  M^2(s;\gamma)}-\alpha(\gamma) s\Bigr] ds = C(\gamma) \exp \Bigl[\frac{-s^2
e^{(\gamma-1)s}}{2M^2} - \alpha(\gamma)s\Bigr] ds \label{eq:PDFgne1}
\end{equation}
where $C(\gamma)$ is a normalizing constant such that
$\int_{-\infty}^{+\infty} P(s;\gamma)ds=1$.
The parameter $\alpha (\gamma) $ is again determined by the constraint of
mass conservation stating that the mean value of the density should
be $1$ : $\int_{-\infty}^{+\infty} e^s P(s;\gamma)ds=1$.
Note that in the absence of cutoffs, the convergence of the integrals
would require $\alpha> 1$ for $\gamma<1$ and $\alpha<0$ for
$\gamma>1$.
This functional form of the PDF immediately allows to make  a few
predictions. For $\gamma<1$, $\hat M(s;\gamma)$ grows exponentially
with $s$ for $s_{-}<s<s_{+}$ and as a consequence the PDF
for $0<s<\hat M(s;\gamma)$ is dominated by the power-law (in $\rho$) behavior
$P(s;\gamma)\sim e^{-\alpha (\gamma )s}$, while the Gaussian-like
decay will again 
dominate for $s>\hat M(s;\gamma)$. For $s<0$, the local
turbulent Mach number decreases as $s$ decreases and we expect a
drop off of the PDF more rapid than when $\gamma =1$. 
The behavior is exactly opposite when
$\gamma>1$. 
This prediction can be verified by looking at Fig.\
\ref{pdfs_vs_gamma_M3} displaying 
the PDF of $s$ for $\gamma=0.3$ and $\gamma=1.7$ at $M=3$.

It is now interesting to relate the PDF for a certain value
of $\gamma$ to that obtained for $2-\gamma$.
Writing the condition $\langle \rho\rangle =1$, we get
\begin{equation}
\int_{-\infty}^{\infty} \exp\Bigl(-\frac{s^2}{2\hat
  M^2(s;\gamma)}+(1-\alpha(\gamma)) s\Bigr)ds =
\int_{-\infty}^{\infty} \exp\Bigl(-\frac{s^2}{2\hat
  M^2(s;\gamma)}-\alpha(\gamma) s\Bigr)ds \label{eq:int1}
\end{equation}
while the same condition for $2-\gamma$ reads, after making the
substitution $s\rightarrow -s$ in the integrals
\begin{equation}
\int_{-\infty}^{\infty} \exp\bigl(-\frac{s^2}{2\hat
  M^2(-s;2-\gamma)}-(1-\alpha(2-\gamma)) s\bigr)ds =
\int_{-\infty}^{\infty} \exp\bigl(-\frac{s^2}{2\hat
  M^2(-s;2-\gamma)}+\alpha(2-\gamma) s\bigr)ds\label{eq:int2}.
\end{equation}

For $s_{-}<s<s_{+}$, the functions 
$\hat M(s;\gamma)$ and  $\hat M(-s;2-\gamma)$ are identical.
If the cutoffs $s_{+}$ and $s_{-}$ occur at large enough values,
i.e. when the local Mach number is either very large or very small,
the contributions in the integrals 
of the two terms involving these two quantities will be very close
and, by inspection of 
eqs. (\ref{eq:int1}) and (\ref{eq:int2}) we get 
\begin{equation}
\alpha(2-\gamma)=1-\alpha(\gamma) \label{eq:symalpha}.
\end{equation}
This relation is exact when $\gamma =1$ 
since $\hat M(s;1)=\tilde
M$ is independent of $s$, allowing to recover the result $\alpha
(1)=\frac{1}{2}$. Note also that for large enough $M$, a case where
eq. (\ref{eq:symalpha}) holds, the symmetry $s\rightarrow -s$ is not
possible but must include a translation in the $s$ domain to account
for mass conservation.

Relation  (\ref{eq:symalpha}) is verified numerically with a
reasonable precision for 
the runs at the highest Mach numbers. For example, when $M=6$, the slope of the
power law is  $-1.2$ (i.e. $\alpha=1.2$) for
$\gamma=0.5$ while we have $\alpha=-0.28$ for $\gamma=1.5$ (see Fig.\
\ref{pdfs_vs_gamma_M6}). For smaller values of $M$, the absolute values of
the slopes are 
closer to each other, a feature due to the different cutoffs for
negative and positive values of $s$ (see Fig.\ \ref{pdfs_vs_gamma_M3}
for $M=3$ and $\gamma=0.3$ and $0.7$).
Note that the shape of the PDF for $M=6$, $\gamma=1.5$ presents a
steeper slope for values of $s$ slightly smaller than that of the
maximum. This feature can  also be
reproduced with this simple model, as can be seen on Fig.\ \ref{theo_pdf}
which displays the PDF obtained from eq. (\ref{eq:PDFgne1}) for $\alpha=
0.28$, $\gamma=1.5$, $\tilde M(0)=1.2$ and values of the cutoffs at
$\tilde M= 10$ for $s<0$ and $\tilde M=0.1$ for $s>0$.

\subsection{The case $\gamma =0$, i.e. Burgers' equation}

An interesting problem concerns the high Mach number limit.
It is often suggested that when $M$ is very large, the dynamics
of compressible flows should be analogous to that prescribed by the Burgers
equation. While this may be true for the velocity field, our results
prove that it cannot be the case for the density.
Indeed, we find that whatever the value of $\gamma \not= 0$ and of the Mach
number, there is always a range of densities for which the pressure
cannot be neglected. For that range of densities the PDF has no
power-law tail but presents a more rapid drop-off.
For $\gamma=1$, it turns out that the pressure is never negligible. 
Extrapolating our results, we thus predict that for the Burgers' case
there should be power law tails both for low and high densities.
We thus performed a simulation of the Burgers equation (coupled with
eq. (\ref{eq:baser}) for the density) with the same
parameters as for the previous runs and with $N=6144$. 
The resulting PDF is presented in Fig.\ \ref{pdf_burg}. This plot shows that
indeed the PDF is almost flat for $s<0$, while there is also a power law for
$s>0$, with a negative slope of roughly $0.5$. The cutoff for large
densities is due to the viscous terms, which give a minimum scale for
the width of the shocks, and thus a maximum value for the density
peaks. In the physical domain, we observe the creation of voids ($s$
reaching a value of $-85$ at $t=64$) which occupy most of the
domain, together
with very  high peaks ($s\approx 6.5$). The number of peaks decreases
during the simulation while the density in the voids decreases
exponentially in time. 
The forcing is unable to break the peaks because it acts at large
scales, while the the density fluctuations become as narrow as allowed
by viscosity.
 This PDF has to be contrasted with the one obtained in
\cite{GK93}, for which the Reynolds number was low and the simulation
decaying. In that case the power law at high densities 
was obtained but the PDF presents a sharp drop off for low densities. 
Two-dimensional decaying simulations of the Burgers equation are also
presented in \cite{SVCP97} for moderate Reynolds numbers. The plateau of the
PDF at low values of the density is also obtained, while the power law for
$s>0$ is not as clear. Burgers simulations for the decaying infinite Reynolds
number case are presented in \cite{VDFN94}. In that case the PDF is
calculated for the cumulated mass function and not for the density,
which is not defined after the first shock formation. A  power law is
found, which extends to $s=-\infty$ and connects to an exponential decay for
$s\rightarrow +\infty$. Note that an exponential PDF for the density
was predicted in \cite{TK72} on the basis of a model which treats shocks
as completely inelastic particles.
We can thus conclude this section by saying
that the Burgers case is truly a singular limit, 
which cannot be reached as the high Mach number limit of a polytropic
gas, with $\gamma\ne 0$. 

\section{Discussion}\label{conclusions}

\subsection{Effects of the forcing}

The study presented in this paper has been performed for a single
choice of the forcing and of the Reynolds number. While the variation
with the latter parameter can be trivially extrapolated, we cannot a
priori be sure that our results are independent of the type of
forcing. We have performed decay runs and observed that 
the behavior of $\sigma_s$ vs. $\tilde M$ is still the same as
in the forced case. The PDFs however  cannot be computed on a single snapshot
due to the poor statistics and cannot be integrated in time since 
the Mach number changes by roughly one or two orders of magnitude during the
run. We have also performed a run at $\gamma=1$ with a forcing of the form
$f/\rho$ in eq. (\ref{eq:baseu}). In that case the density PDF is not a
lognormal anymore but presents a power law tail for low
densities (not shown). This can be attributed to the fact that the flow is 
stirred more vigorously at low densities so that the effective Mach
number indeed increases as $\rho$ decreases.
We nevertheless think that our results can be extrapolated to an
unforced situation, at a given time, and possibly also to the
multi-dimensional case. 
Note that the Mach numbers we have explored in this paper would
correspond to even higher Mach numbers in the multi-dimensional case
since in that case only a fraction of the total kinetic energy populates the
compressible modes.  

\subsection{Summary}

We have presented an investigation of the density PDFs of a randomly
accelerated polytropic gas for different values of the polytropic
index and of the Mach number. We have suggested a simple model in
which the density field is everywhere constructed by a random
succession of jumps \cite{V94}. When the flow is isothermal
($\gamma=1$), the jumps are independent of the initial density, and
have always the same probability distribution. Expressed with the
variable  $s\equiv \ln
\rho$ the jumps are additive, and by the Central Limit Theorem are
expected to have a Gaussian PDF, or a lognormal in $\rho$. 

An analysis of the expected $s$ increments in the weak and strong
shock cases, as well as those due to expansion waves, suggested that
the variance $\sigma_s^2$ should scale as the mean square turbulent
Mach number $\tilde M^2 $. Moreover, because of mass conservation, the
peak of the distribution $s_o$ is related to the variance by
$s_o=-\frac{1}{2}\sigma_s^2$. These predictions were verified in 1D
simulations of compressible turbulence. Previous claims that it is the
{\it density} variance $\sigma_{\rho}^2$ that should scale as $\tilde
M^2$ \cite{PNJ97} might have been misled by lower effective
Mach numbers than those 
achieved in the present simulations, in which all of the kinetic
energy is in compressible modes thanks to the one-dimensionality.

When $\gamma \not=1$, the density jumps are not independent of the local
density anymore, and the shape of the PDF should change. Observing that a
renormalization of the Mach parameter (eq.\ (\ref{eq:baseu}))
$M\rightarrow  M(s;\gamma)= Me^{\frac{1-\gamma}{2}s}$ restores the form
of the equations for the case $\gamma=1$, we proposed the ansatz that
the PDF may still be described by the same functional form as in the
case $\gamma=1$, but substituting $M$ by $M(s;\gamma)$. This
prediction is confirmed by the numerical simulations, giving PDFs
which are qualitatively in very good agreement with the model PDF,
eq.\ (\ref{eq:PDFgne1}). The result is that the PDF asymptotically
approaches a power law on the side where $(\gamma-1)s<0$,
while it decays faster than lognormally on the other side.

Upon the replacements $\gamma \rightarrow (2-\gamma)$ and
$\rho \rightarrow 1/\rho$ we find, using the condition of 
mass conservation, that the slope $\alpha$ of the power law for a
given value of $\gamma$ is related to its value at $2-\gamma$ by
eq.\ (\ref{eq:symalpha}) in the large Mach number limit. These results
are also confirmed by the numerical simulations, which exhibit a power
law at $s>0$ when $\gamma<1$ and at $s<0$ when $\gamma>1$, with slopes
which are roughly related by eq.\ (\ref{eq:symalpha}), with better accuracy
at large Mach numbers.

Finally, on the basis of these results, we suggested that the Burgers
case should develop a power law PDF at both large and small densities,
since in this case there is no pressure on either side. This result
was again confirmed by a simulation of a Burgers flow. 

We shall conclude this paper by pointing out that the non-uniqueness
of the infinite Mach number limit might have important consequences
for astrophysical applications, such as in cosmology. 
The so-called Zeldovich \cite{Z70} approximation is indeed based on
the Burgers' equation which, at the light of the present work,
appears as a questionable model of highly compressible
flows. This point will be addressed in future work.

\begin{acknowledgements}

We thankfully acknowledge financial support from UNAM/CRAY grant
SC-008397 and UNAM/DGAPA grant IN105295 to
E.\ V.-S., and a joint CONACYT-CNRS grant. This work also  benefited
from partial support from the Programme National du CNRS ``Physique
Chimie du Milieu Interstellaire''.

\end{acknowledgements}

%
%
\vskip 1truecm
\centerline{\bf Figure captions}

\begin{figure}
\caption{(Top) Variance of $s=\ln \rho$  vs.\ the mean square Mach number
$M_{\rm rms}^2=\tilde M^2$ for various simulations with $\gamma=1$ and
$M=0.5$, 1, 2, 
3, 4.5, and 6. Every point in this plot gives the variance and
$\tilde M$ for sets of 100 subsequent outputs (10 time units) of any
given simulation. The simulations were typically run for 150 time
units. (Bottom) Variance of $\rho$ vs.\ $\tilde M$.}
\label{sig_vs_M_g1}
\end{figure}

\begin{figure}
\caption{Most probable value of $s$ vs the variance of $s$,
$\sigma_s^2$, for the runs in Fig.\ \ref{sig_vs_M_g1}. 
The data points are obtained as in Fig.\ \ref{sig_vs_M_g1}.}
\label{s_vs_sigs_g1}
\end{figure}

\begin{figure}
\caption{Probability density function (PDF) of $s$ for three
simulations with $\gamma=1$ and $M=$ 0.5, 2 and 6. For clarity, these
PDFs have been respectively displaced in the plot by $-2$,
$-1$, and 0 units in the vertical axis. The shift of the peak towards
more negative $s$ values at larger $M$ is real, due to the constraint of mass
conservation. The dashed lines show the best fit with a lognormal to
each PDF.}
\label{pdfs_vs_M}
\end{figure}

\begin{figure}
\caption{PDF of the velocity derivative for a run with $\gamma=1$ and $M=6$.}
\label{pdf_ux}
\end{figure}

\begin{figure}
\caption{a) (Top) Density field of a run with $\gamma=0.5$ and $M=10$ at time
$t=34.65$ Note the very thin density peaks and the shallow density
minima. b) (Bottom) Density field of a run with $\gamma=1.5$ and $M=6$ at
$t=50.5$. Note that the density maxima are now much shorter, while the
density minima (voids) become much deeper. They are also much wider
than the peaks in the $\gamma=0.5$ case because of  mass conservation.}
\label{rho_fields}
\end{figure}

\begin{figure}
\caption{Standard deviation of $s$ vs.\ the mean value of $s$ over
subregions of size 1/3 of the 
integration domain for two runs with (top) $\gamma=0.5$ and (bottom)
$\gamma=1.5$. Note the inverse trends between the two runs and the
saturation of $\sigma_s$ at large values of $|\langle s \rangle|$,
especially noticeable in the case $\gamma=1.5$.}
\label{sig_vs_avg}
\end{figure}

\begin{figure}
\caption{Variance of $s$ (top) and of $\rho$ (bottom) vs.\ the mean
square Mach number for 6 runs with $\gamma=0.5$ and $M=0.5$, 2,
3, 4.5, 6 and 10. Note that $\sigma_s^2$ increases more slowly than
$M_{\rm rms}^2$ because only one side
($s>0$) of the density PDF is unimpeded by the
pressure. Instead, $\sigma_\rho^2$ increases more rapidly than $M_{\rm
rms}^2$ because such fluctuations in $s$ imply very large fluctuations
in $\rho$.}
\label{sig_vs_M_g05}
\end{figure}

\begin{figure}
\caption{Variance of $s$ (top) and of $\rho$ (bottom) vs.\ the mean
square Mach number for 6 runs with $\gamma=1.5$ and $M=0.5$, 2,
3, 4.5 and 6. Again (compare to Fig.\ \ref{sig_vs_M_g05}),
$\sigma_s^2$ increases more slowly than $M_{\rm rms}^2$ because
only one side $s<0$ of the PDF is unimpeded by the pressure. However,
in this case also $\sigma_\rho^2$ increases more slowly than $M_{\rm
rms}^2$, because the density fluctuations are bounded by zero, not
being able to contribute much to the variance of $\rho$.}
\label{sig_vs_M_g15}
\end{figure}

\begin{figure}
\caption{PDFs of $s$ for two simulations with $M=3$ and $\gamma=0.3$
(top) and $\gamma=1.7$ (bottom). For $\gamma=0.3$ the power-law regime
appears at large densities, while for $\gamma=1.7$ it appears at small
densities.}
\label{pdfs_vs_gamma_M3}
\end{figure}

\begin{figure}
\caption{PDFs of $s$ for two simulations with $M=6$ and $\gamma=0.5$
(top) and $\gamma=1.5$ (bottom). Note that at this Mach number, the
power-law regime for the $\gamma=1.5$ case appears removed from the
peak of the distribution, mediated by a regime with a steeper slope.}
\label{pdfs_vs_gamma_M6}
\end{figure}

\begin{figure}
\caption{The theoretical PDF given by eq.\
(\ref{eq:PDFgne1}) for $\alpha=0.28$, $\gamma=1.5$, $\tilde M(0)=1.2$
and values of the cutoffs at $\tilde M= 10$ for $s<0$ and $\tilde
M=0.1$ for $s>0$. Compare to Fig.\ \ref{pdfs_vs_gamma_M6}.}
\label{theo_pdf}
\end{figure}

\begin{figure}
\caption{PDF of $s$ for a Burgers run. Note the nearly flat slope at
negative-$s$ values.}
\label{pdf_burg}
\end{figure}

%
%


\begin{references}
\bibitem{footnote} Note that $\gamma$ is here simply a parameter,
and need not be related to the heat capacity ratio of the gas.
\end{references}
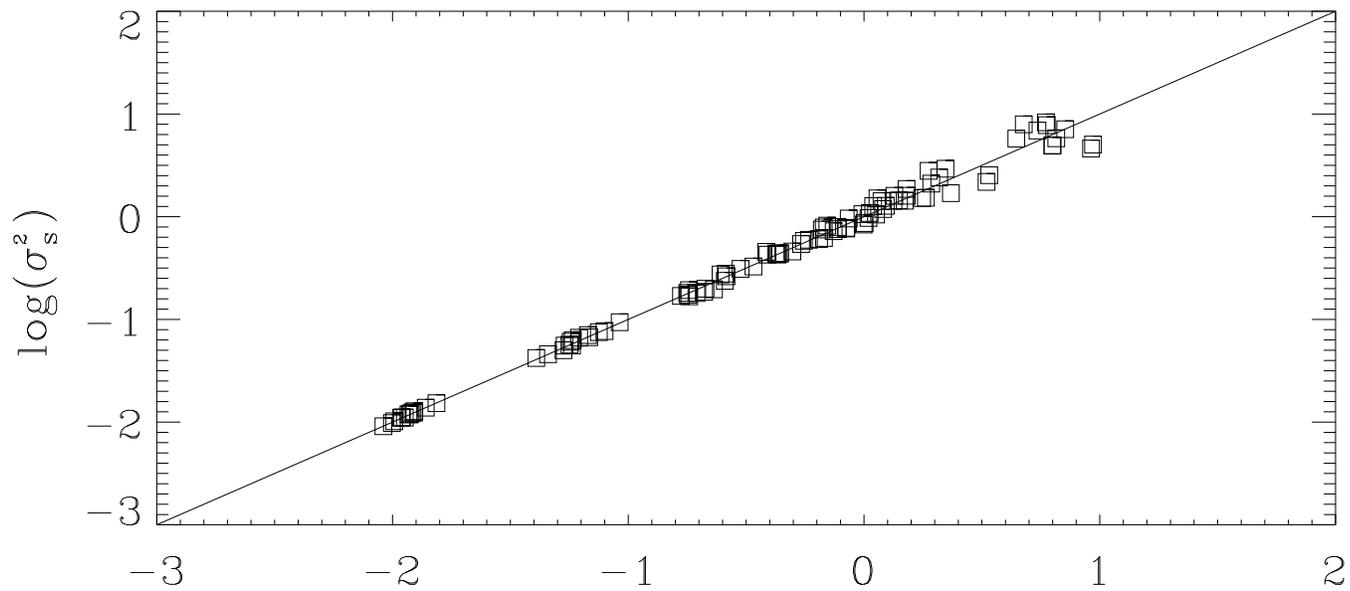
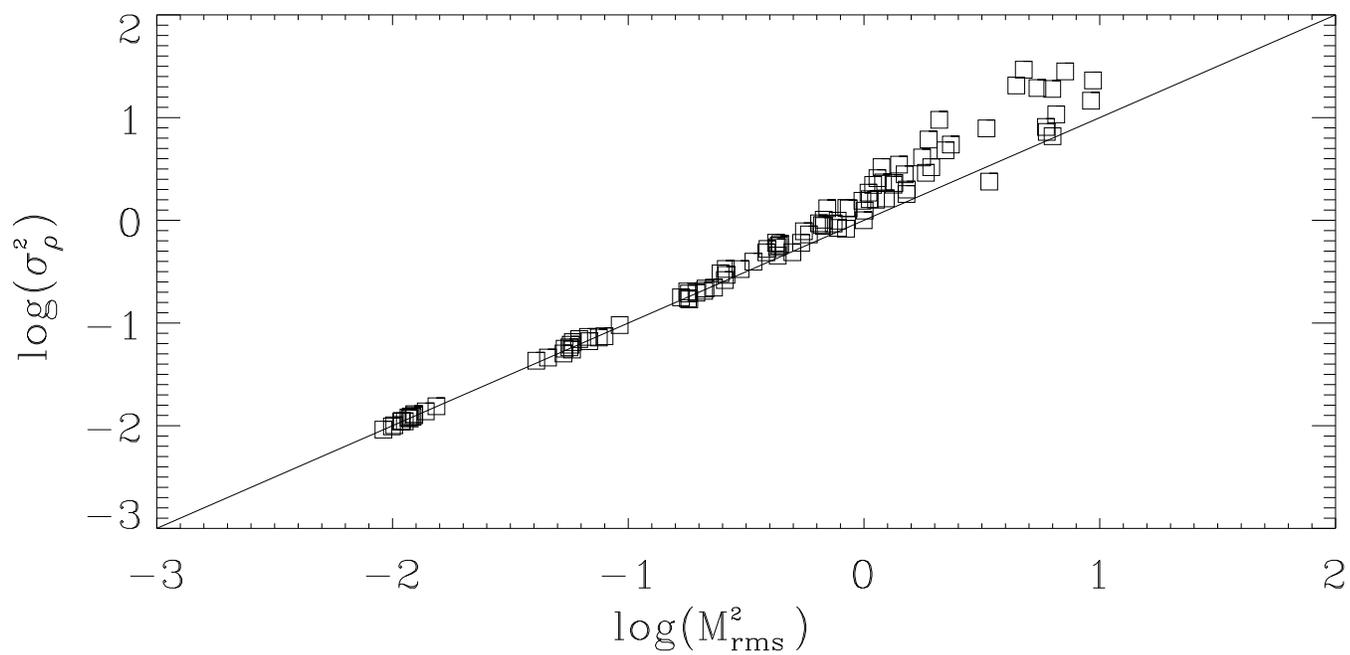
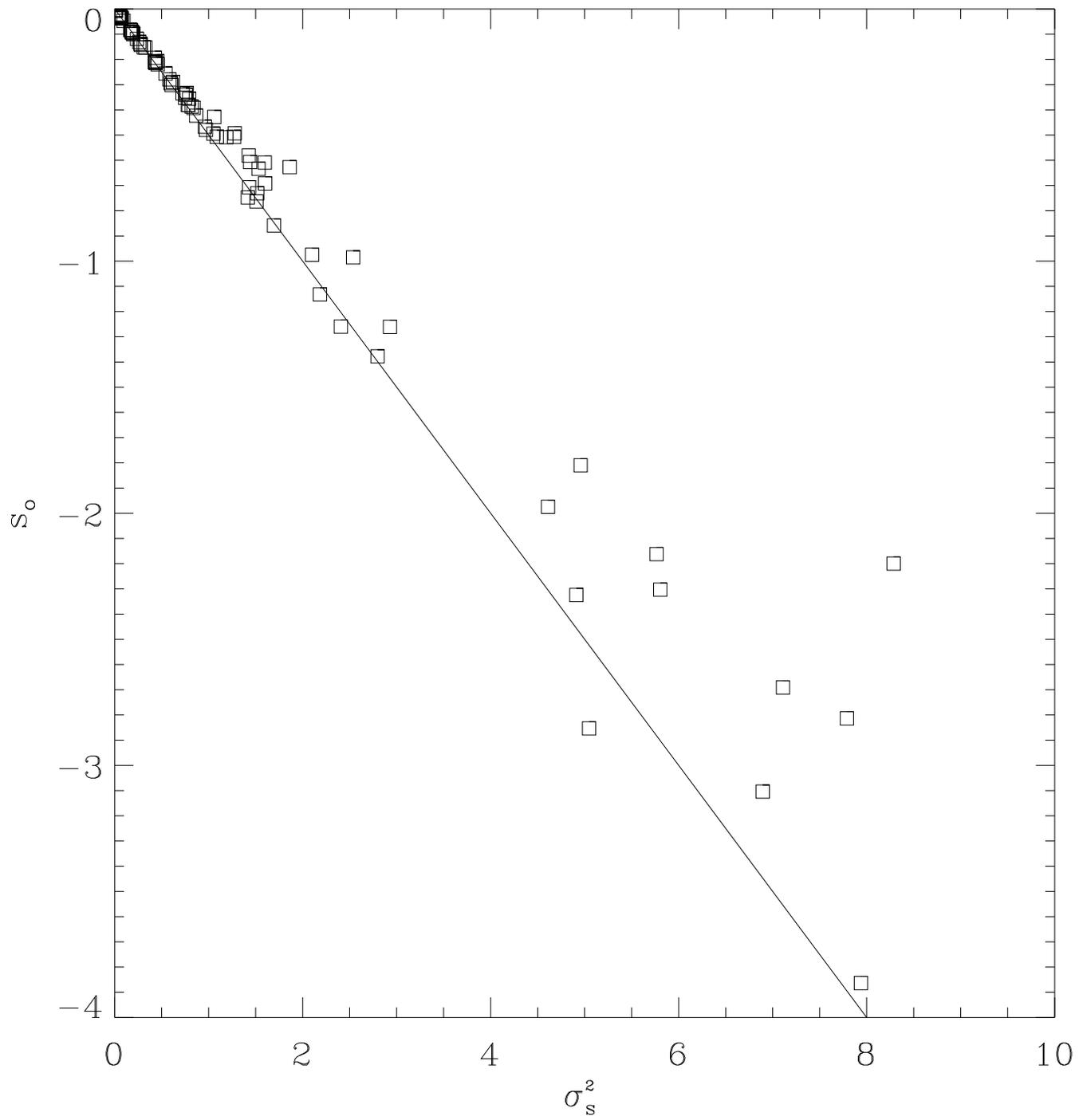
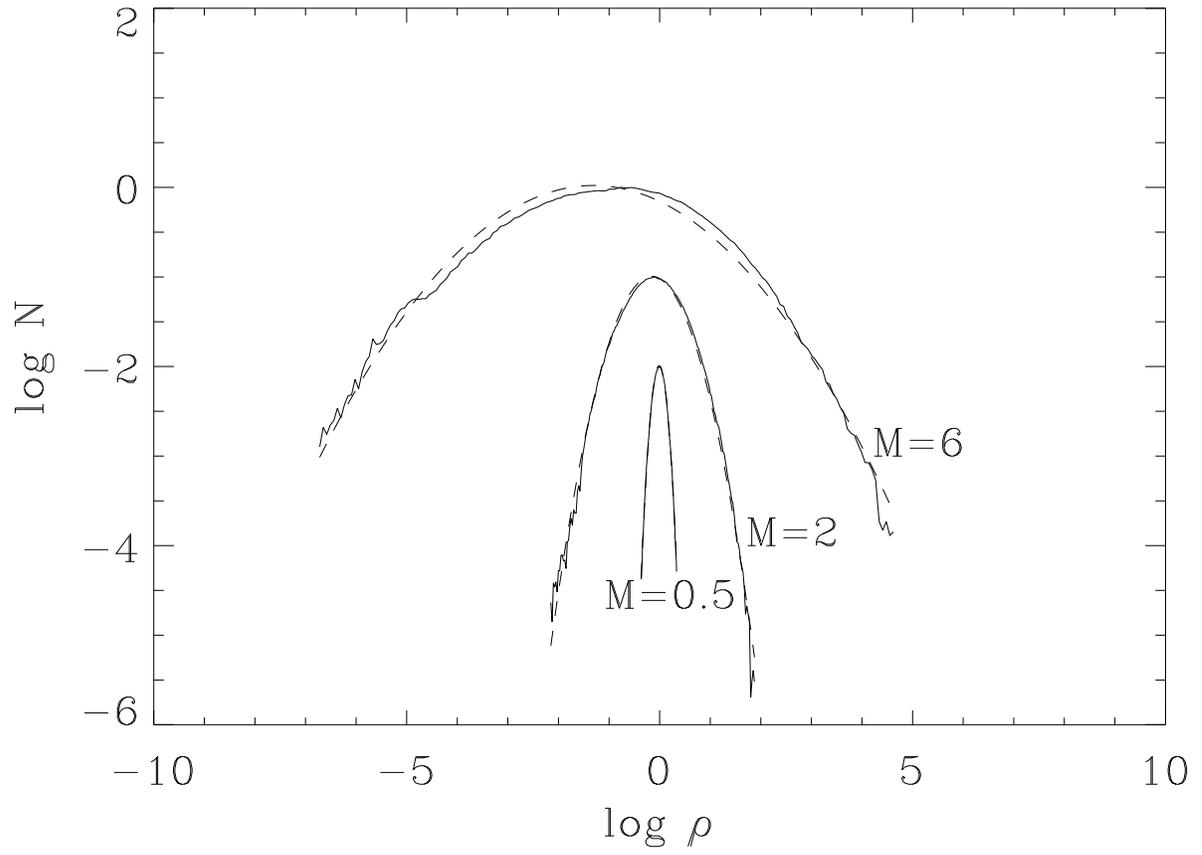
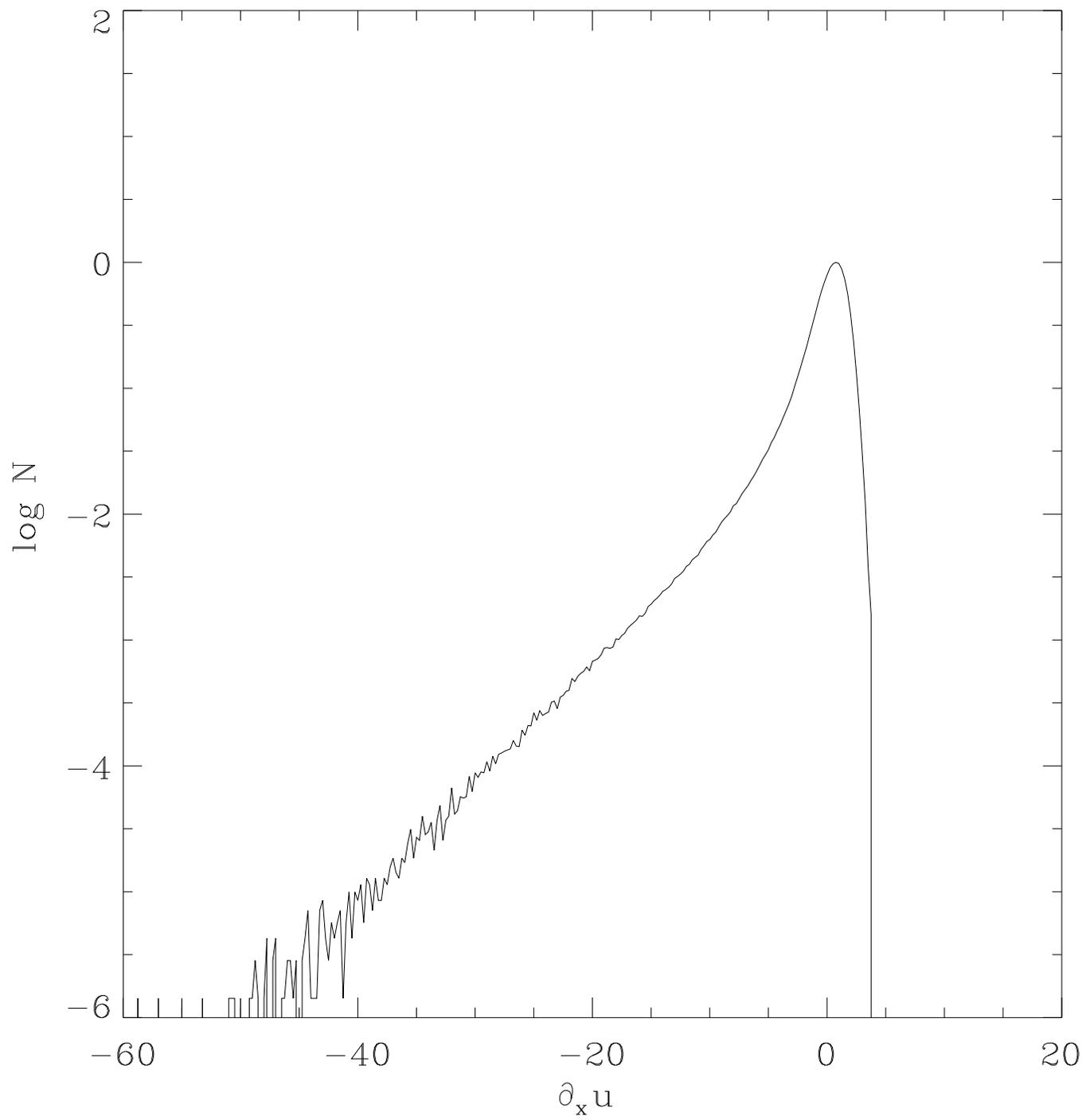
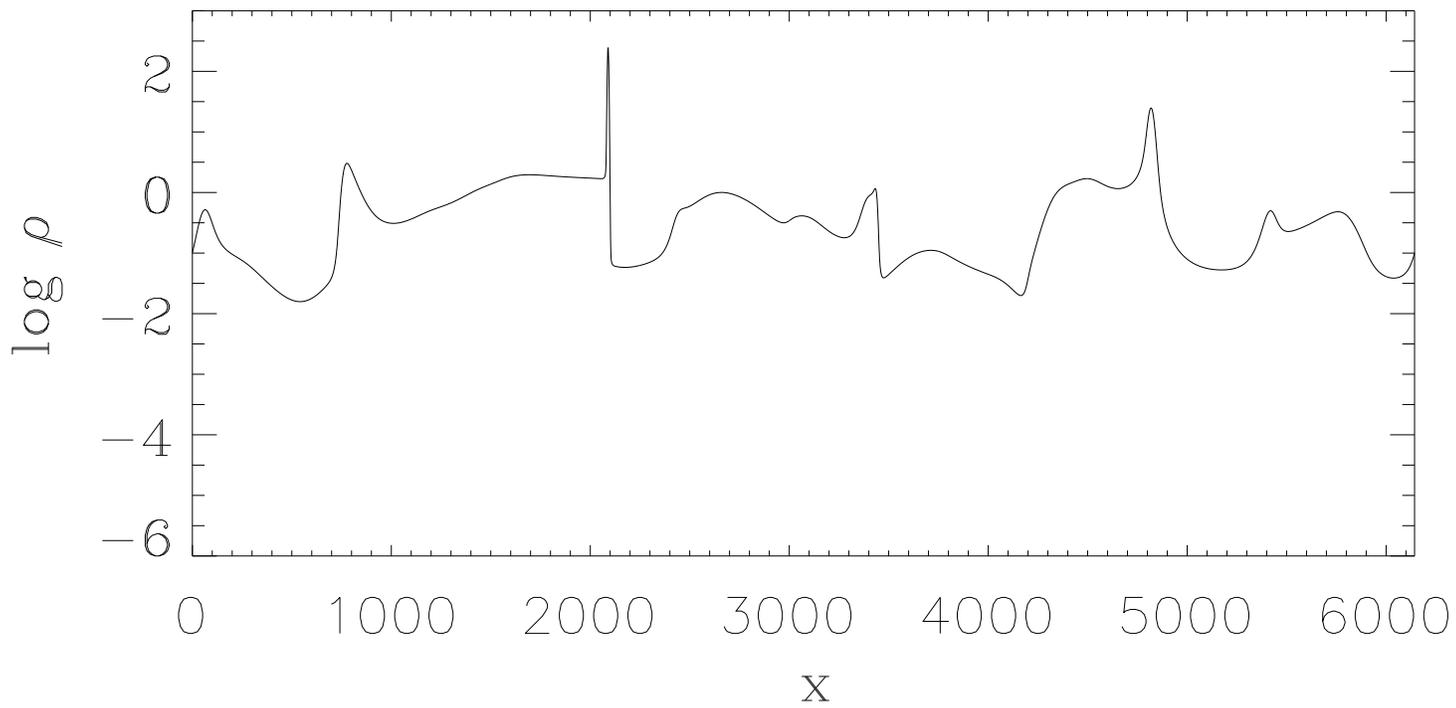
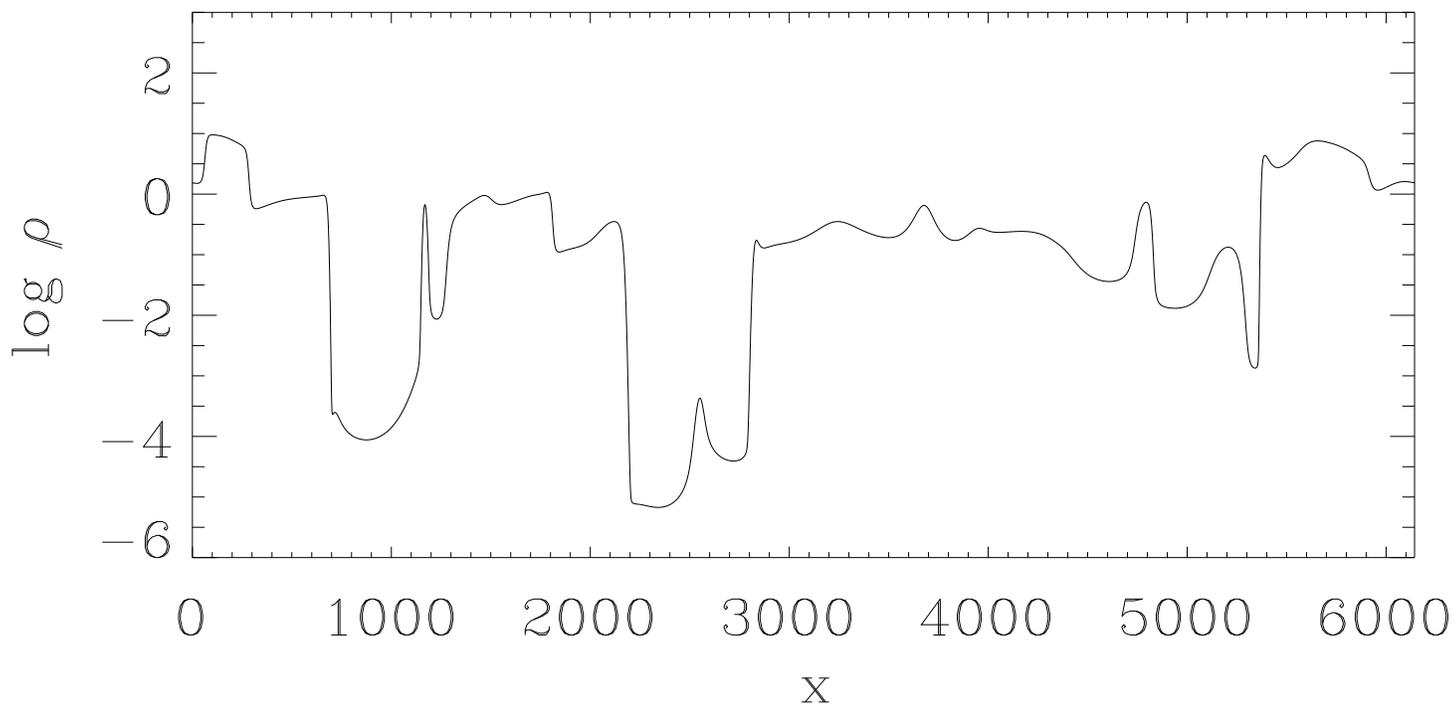
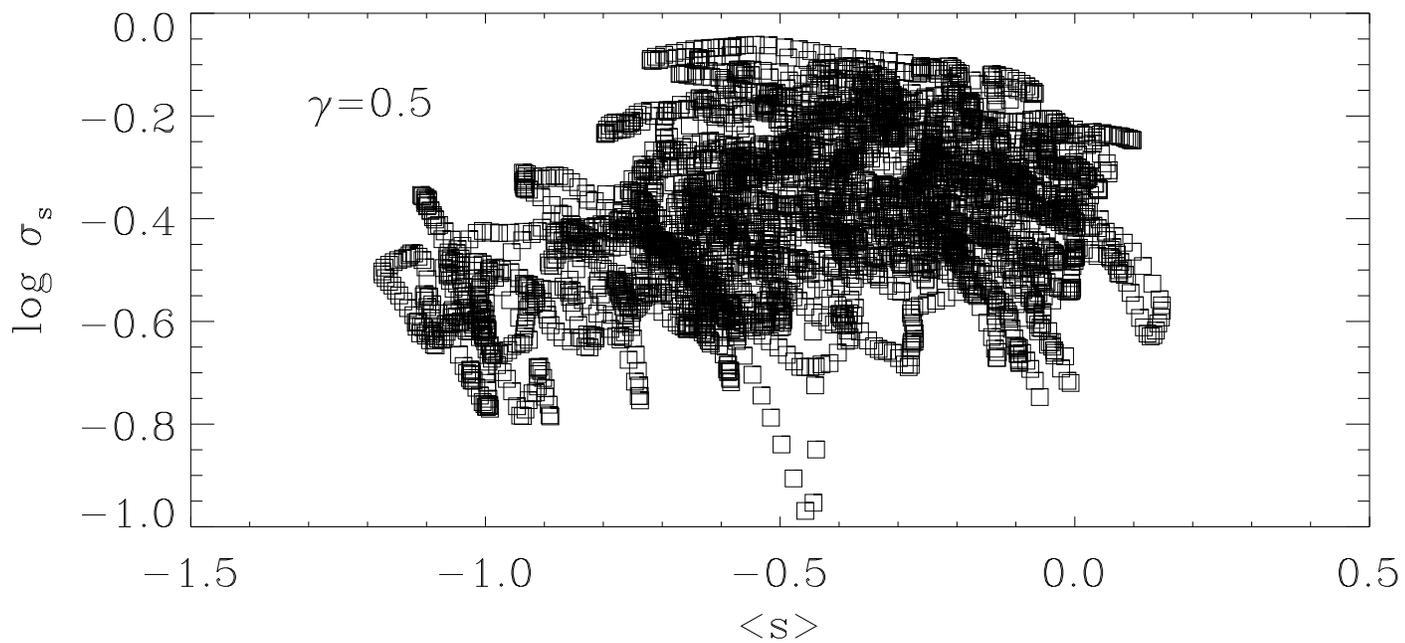
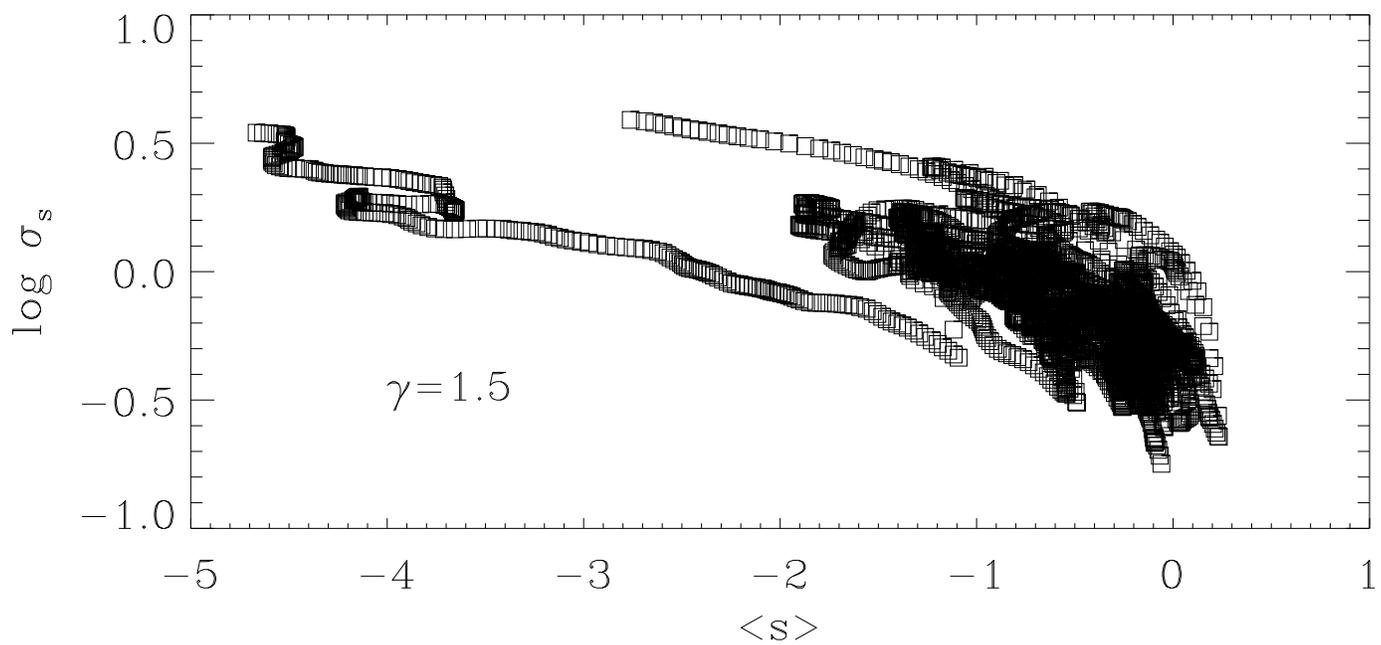
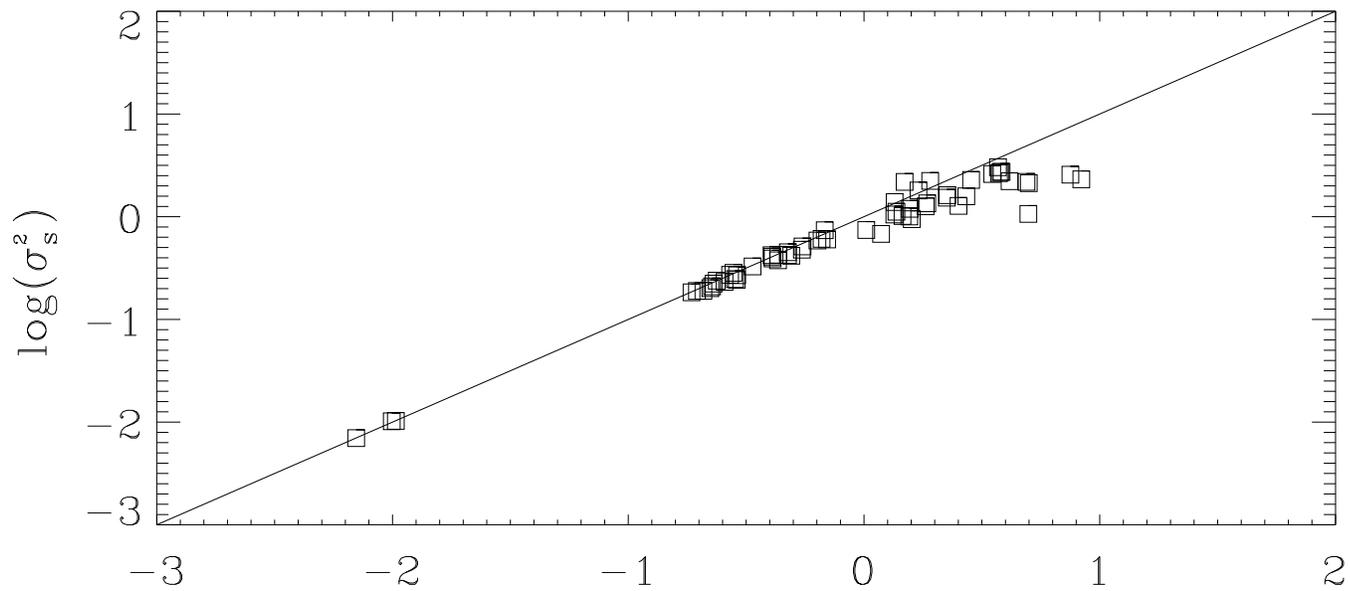
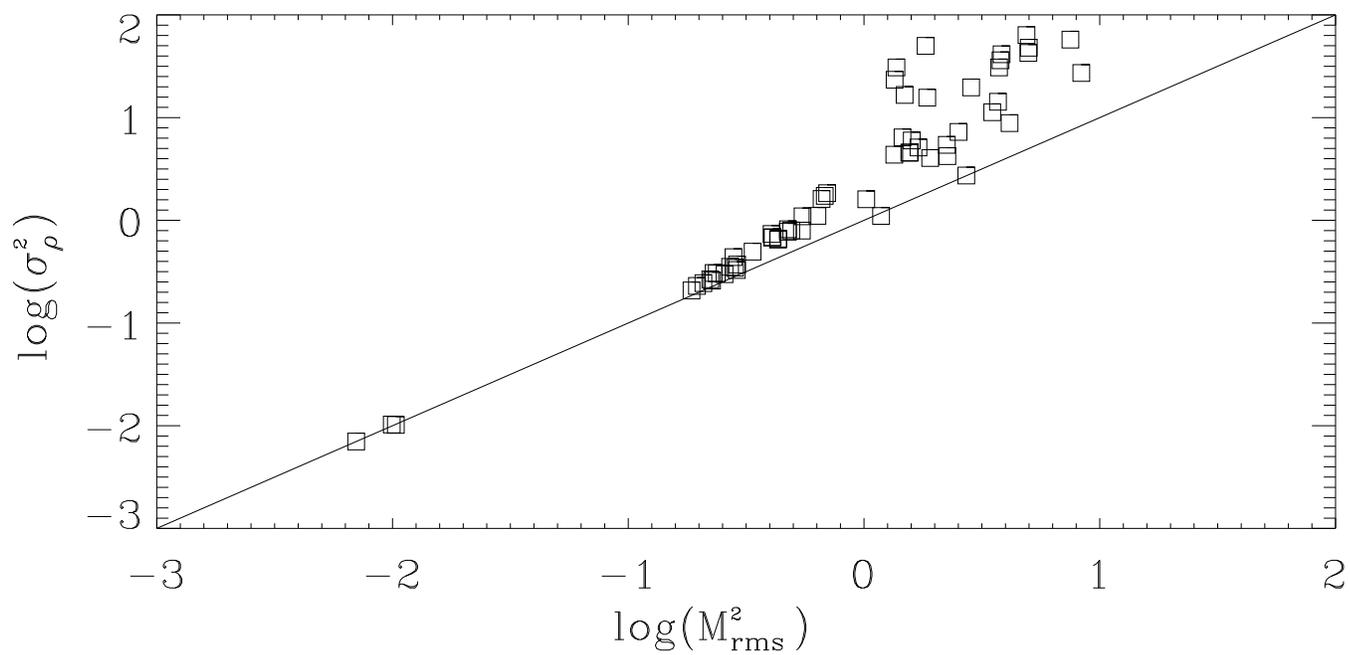
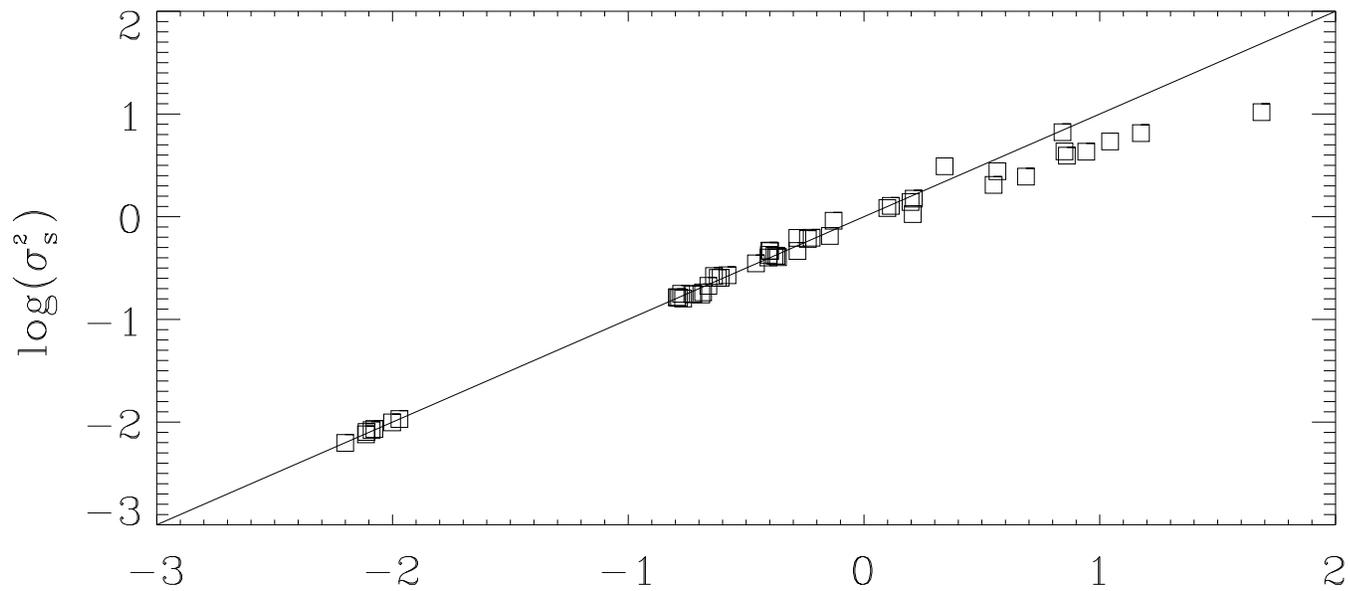
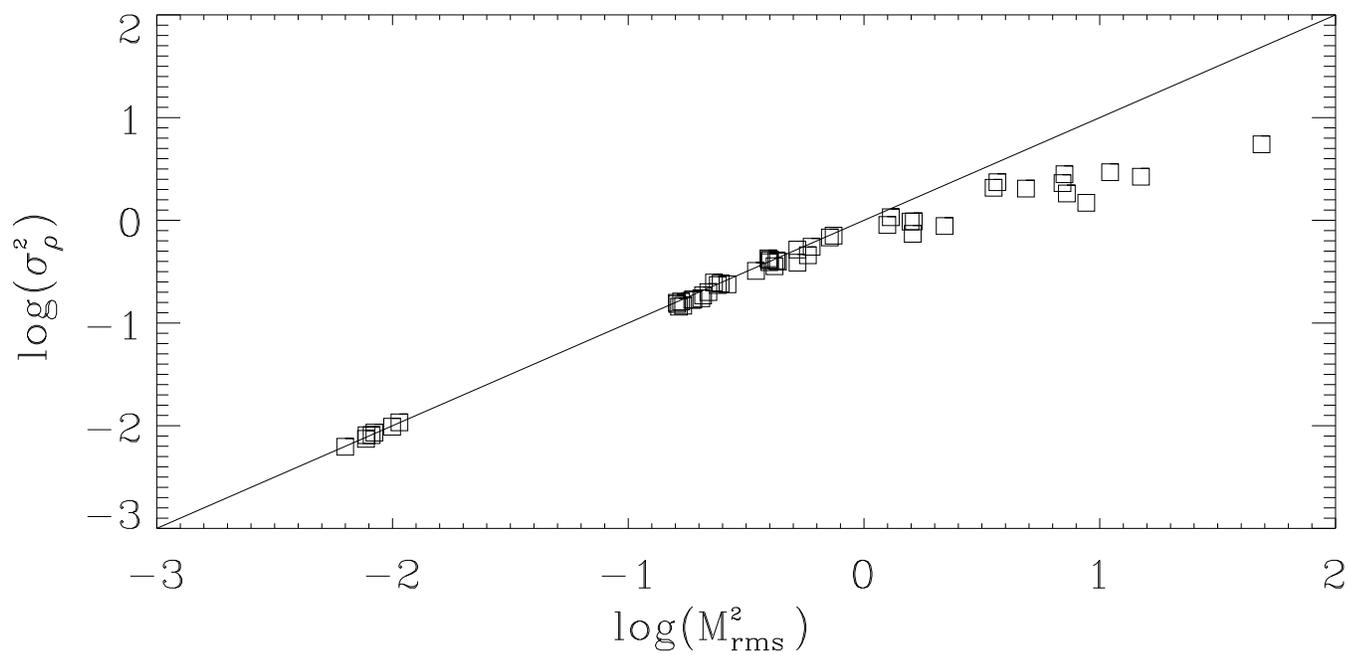
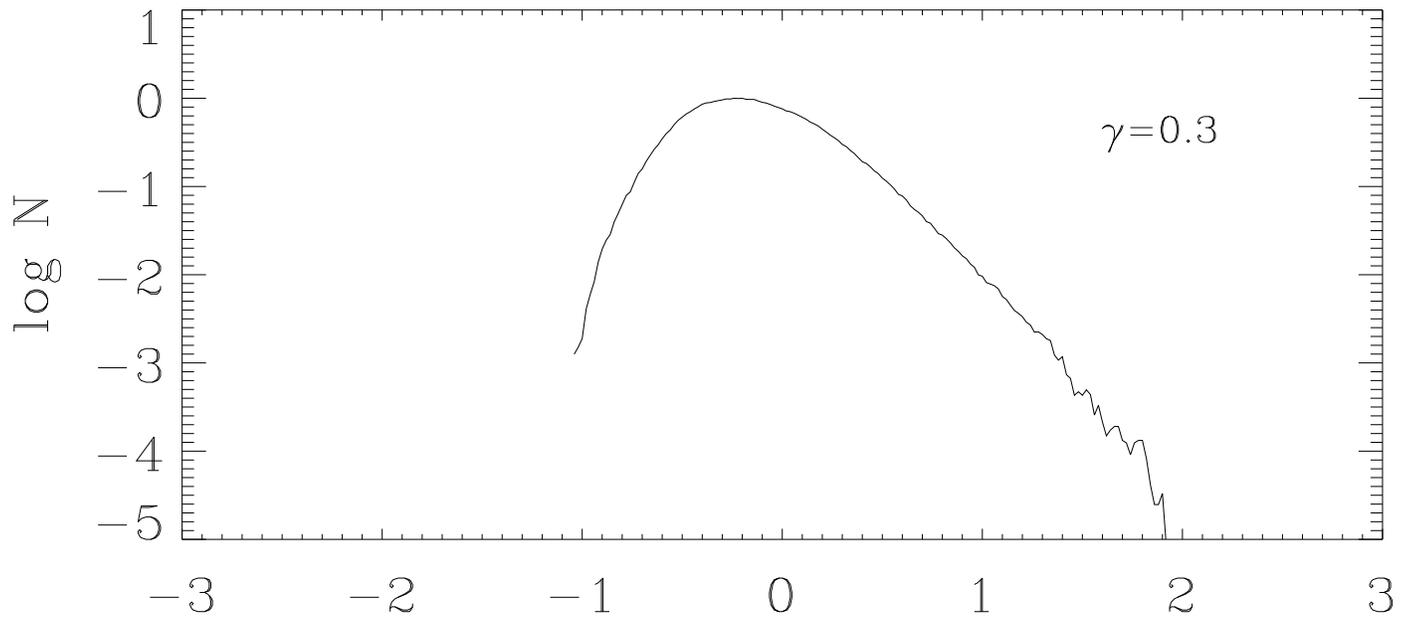
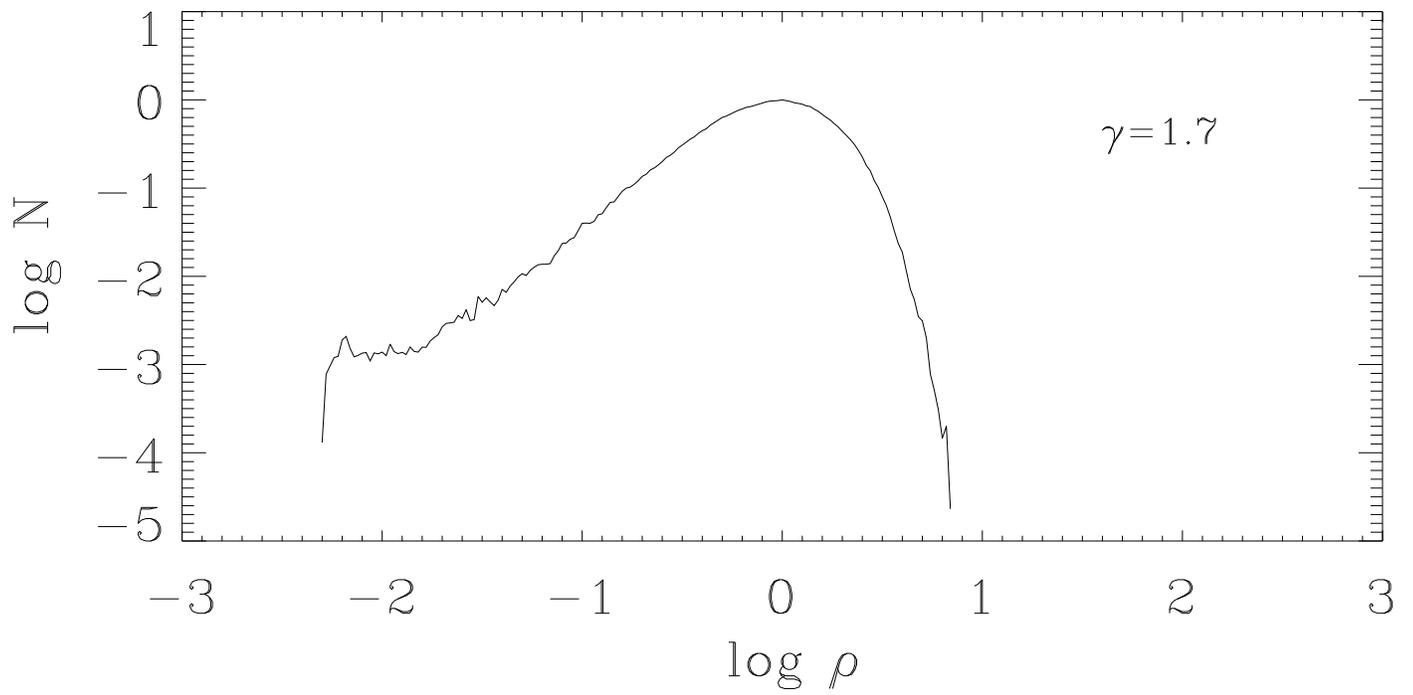
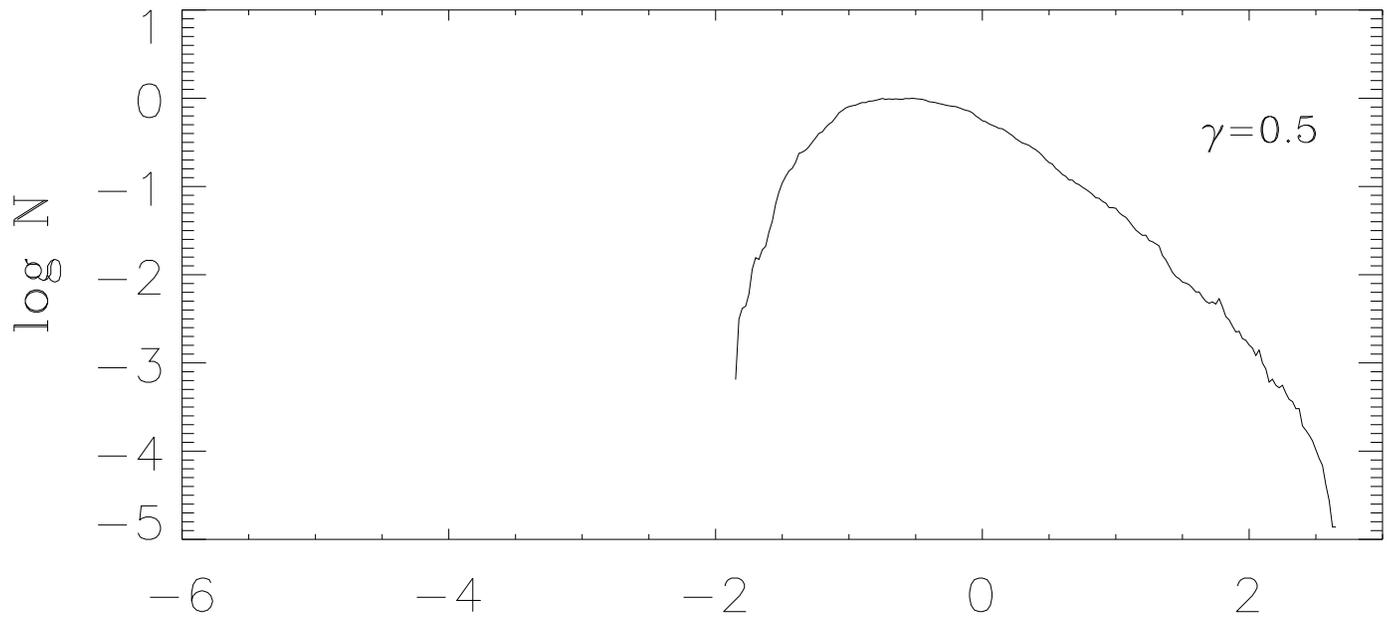
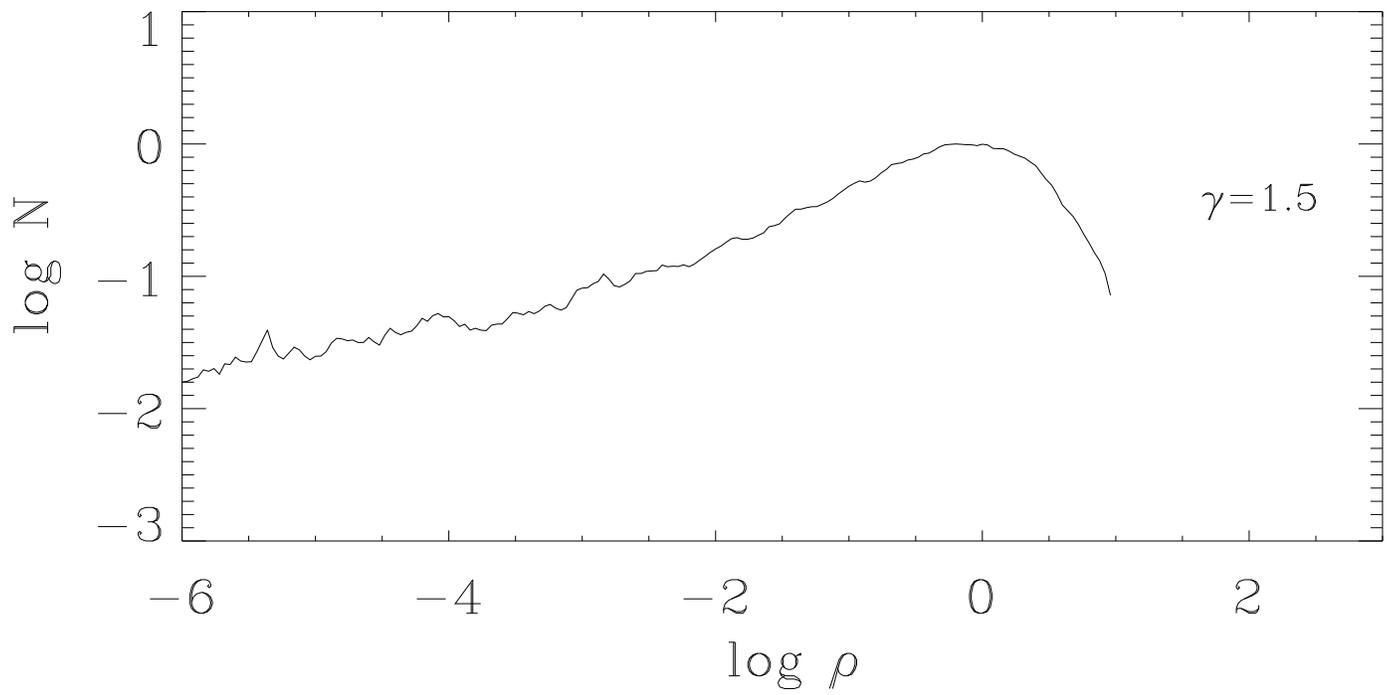
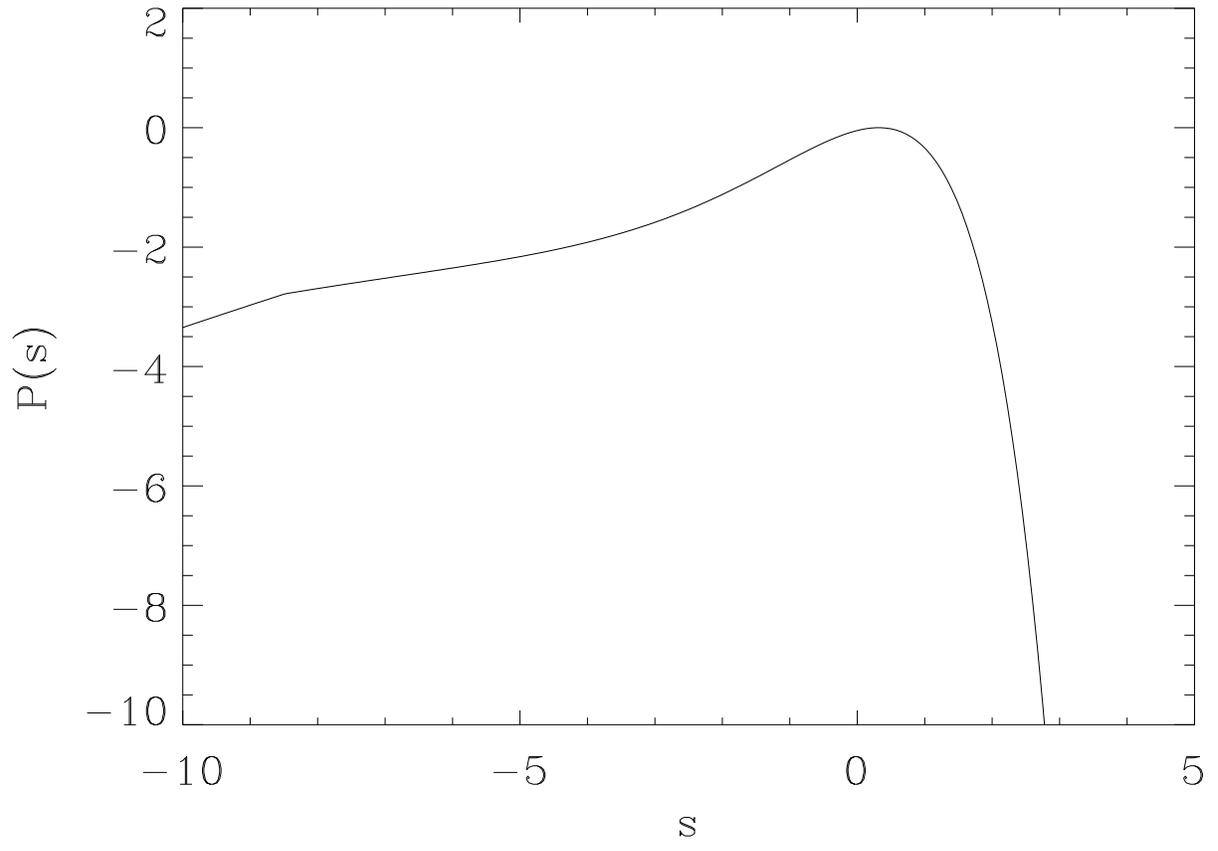
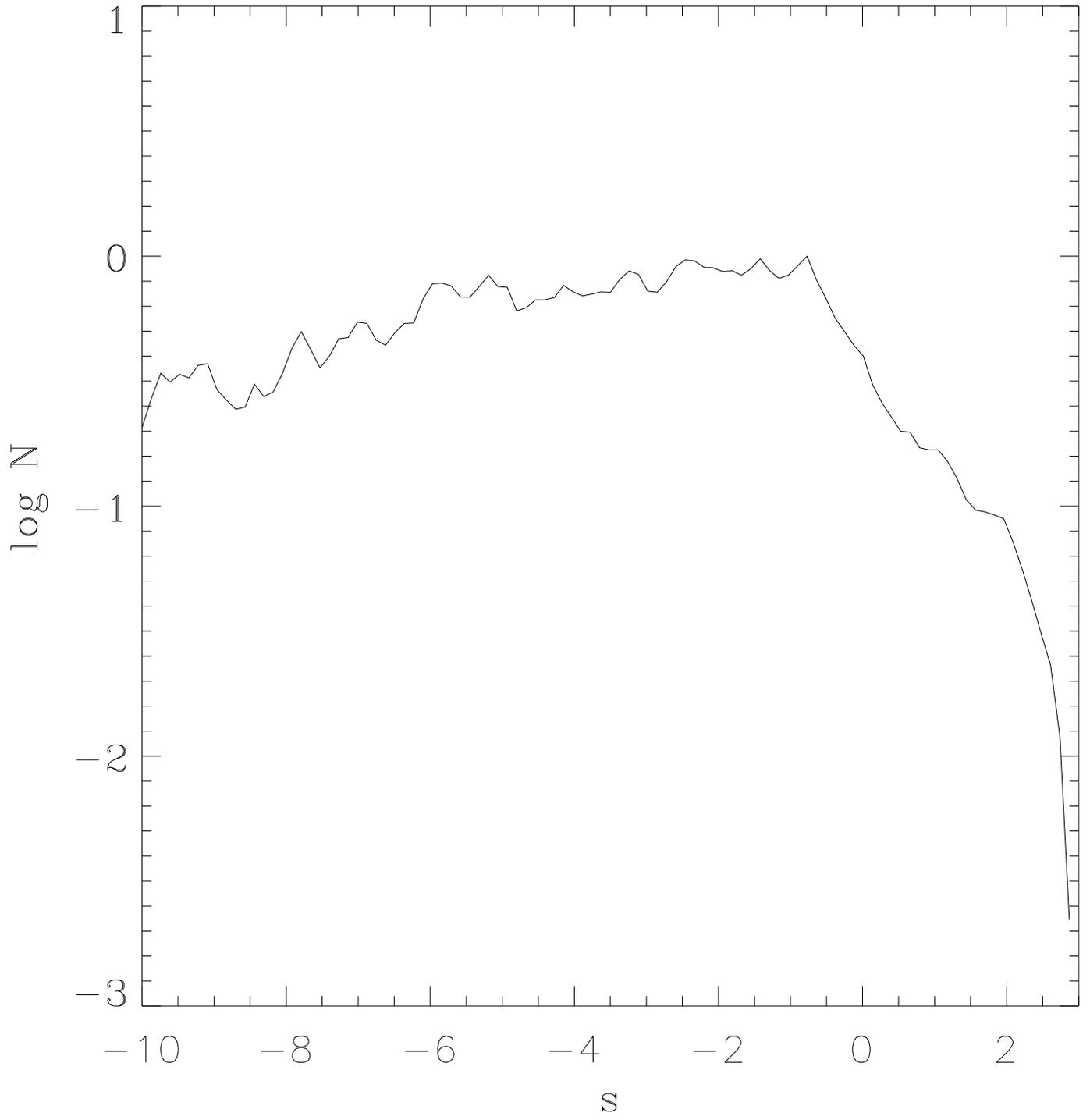
\end{document}